\documentclass[prd,twocolumn,showpcs,amsmath,amssymb,nofootinbib,preprintnumbers,balancelastpage]{revtex4}
\pdfoutput=1
\usepackage{epsfig}
\usepackage{amsmath}
\usepackage{bm}
\usepackage{times}
\usepackage{graphicx}
\usepackage{color}
\usepackage{slashed}
\usepackage{graphicx}
\usepackage{amsmath}

\def\bea{\begin{eqnarray}}
\def\eea{\end{eqnarray}}
\def\ba{\begin{eqnarray}}
\def\ea{\end{eqnarray}}
\def\be{\begin{equation}}
\def\ee{\end{equation}}

\begin{document}
\preprint{CALT 68-2931}

\title{Phenomenology of scalar leptoquarks}

\author{Jonathan M. Arnold, Bartosz Fornal and Mark B. Wise\\
\textit{California Institute of Technology, Pasadena, CA 91125, USA}\\
}
\date{\today}

\begin{abstract}
We study the simplest renormalizable scalar leptoquark models where the standard model is augmented only by one additional scalar representation of $SU(3)\times SU(2)\times U(1)$. The requirement that there be no proton decay from renormalizable interactions singles out two such models, one of which exhibits an unusual top mass enhancement of the $\mu \rightarrow e \gamma$ decay rate. We analyze the phenomenology  of the model with the unusual top mass enhancement of loop level chirality changing charged lepton processes in the light of existing and upcoming experiments. Both of the models that do not allow proton decay from renormalizable interactions have dimension five operators that, even if suppressed by the Planck scale, can give rise to an unacceptably high level of baryon number violation. We discuss symmetries that can forbid these dimension five operators.
\vspace{11mm}
\end{abstract}

\maketitle
\bigskip

\section{Introduction}
Currently, the standard model describes most aspects of nature with remarkable precision. If there is new physics at the multi TeV scale (perhaps associated with the hierarchy puzzle), it is reasonable to expect measurable deviations from the predictions of the standard model  in the flavor sector.  Amongst the experiments with very high reach in the mass scale associated with beyond the standard model physics are those that look for flavor violation in the charged lepton sector through measurements of the processes,  $\mu \rightarrow e \gamma$ \cite{Adam:2011ch} and $\mu \rightarrow e$ conversion \cite{Abrams:2012er,Kuno}, and the search for electric dipole moments of the neutron, proton and electron.

Models with scalar leptoquarks can modify the rates for these processes. Simple models of this type have been studied previously in the literature, including their classification and phenomenology \cite{Buchmuller:1986zs,Davies:1990sc,Davidson:1993qk,Gabrielli:2000te,Benbrik:2008si,Dorsner:2009cu,Benbrik:2010cf,Gonderinger:2010yn}.

Our approach is to first identify the minimal renormalizable scalar leptoquark models containing one single additional representation of $SU(3)\times SU(2)\times U(1)$ and construct the most general renormalizable model without any additional constraints on the couplings apart from the usual ones, i.e., gauge invariance, Poincar\'{e} invariance,  and locality. Given the strong experimental constraints on baryon number violating processes like $p \rightarrow \pi^0 e^+$,  we concentrate only on those scalar leptoquark models which don't have baryon number violation in perturbation theory. Of course there is baryon number violation through non-perturbative quantum effects since it is an anomalous symmetry. But this is a very small effect at zero temperature. Only two models fulfill this requirement.
One of those two models gives a top mass enhanced $\mu \rightarrow e \gamma$ decay rate. We perform an  analysis of the phenomenology of this specific model, including the $\mu \rightarrow e \gamma$ decay rate, $\mu \rightarrow e$ conversion rate, as well as electric dipole moment constraints focussing mostly on the regions of parameter space where the impact of the top quark mass enhancement is most important. For lepton flavor violating processes at higher energies such as $\tau \rightarrow \mu \gamma$, deep inelastic scattering $e+p \rightarrow \mu (\tau) +X$, {\it etc.}, the impact on the phenomenology of the top quark mass enhancement of charged lepton chirality flip is less dramatic and that is why we focus in this paper on low energy processes involving the lightest charged leptons.

There is also an $m_t$ enhancement of the one-loop contribution to the charged lepton mass matrix. We focus on the region of parameter space where this contribution does not necessitate a fine-tuning of parameters.

We also consider the effects of dimension five operators that can cause baryon number violation. We find that the two models without renormalizable baryon number violation can have such operators and, even if the operators are suppressed by the Planck scale, they may (depending on the values of coupling constants and masses) give rise to an unacceptable level of baryon number violation. We discuss a way to forbid these dimension five operators.

\section{Models}

A general classification of renormalizable leptoquark models can be found in \cite{Buchmuller:1986zs,Davies:1990sc}. However, in the spirit of our approach, in which we are interested in models with no proton decay, a more useful list of possible interaction terms between the scalar leptoquarks and fermion bilinears is presented in \cite{Arnold:2012sd}, where those models that have tree-level proton decay are highlighted. The relevant models are listed in Table I below.
\begin{table}[h!]
\begin{center}
    \begin{tabular}{| c | c | c|}
    \hline
       leptoquark & diquark & \ \ \ \ \ $ SU(3) \times SU(2) \times U(1)$ \ \ \ \ \   \\
       \raisebox{0.3ex}[0pt]{couplings} & \raisebox{0.3ex}[0pt]{couplings} & \ \ \ \ \ \raisebox{0.3ex}[0pt]{representation of $X$}\ \ \ \ \   \\ \hline\hline
       $X\bar{Q}e$, $X L \bar{u}$ & $-$ & $\left(3,2, 7/6\right)$  \\ \hline
    $X L\bar{d}$ & $-$ & $\left(3,2, 1/6\right)$  \\ \hline
    $\ \ X\bar{Q}\bar{L}$, $X\bar{u}\bar{e} \ \ $ & $\ \ X Q Q$, $X u d \ \ $ &$\left(3,1, -1/3\right)_{\rm PD}$  \\ \hline
     $X\bar{Q}\bar{L}$ & $X Q Q$& $\left(3,3, -1/3\right)_{\rm PD}$  \\ \hline
    $X\bar{d}\,\bar{e}$ & $X u u$ & $\left(3,1, -4/3\right)_{\rm PD}$  \\ \hline
    \end{tabular}
\end{center}
\caption{\footnotesize{Possible interaction terms between the scalar leptoquarks and fermion bilinears along with the corresponding quantum numbers. Representations labeled with the subscript ``PD'' allow for proton decay via tree-level scalar exchange.}}
\label{table1}
\end{table}

The only two models fulfilling our requirement are $X = \left(3,2, 7/6\right)$ and $X = \left(3,2, 1/6\right)$ .
\\

\noindent $\textbf{Model I:\ \ \ }X = \left(3,2, 7/6\right)$. \\

The Lagrangian for the scalar leptoquark couplings to the fermion bilinears in this model is,
\bea
\mathcal{L} &=& -\lambda_u^{ij} \bar{u}_{R}^i X^T\epsilon L_L^j -\lambda_e^{ij} \bar{e}_{R}^i X^\dagger Q_L^j + {\rm h.c.}\ ,
\eea
where,
\bea
X = \left(
      \begin{array}{c}
        V_\alpha \\
        Y_\alpha \\
      \end{array}
    \right) \ , \ \ \ \epsilon = \left(
                                   \begin{array}{cc}
                                     0 & 1 \\
                                     -1 & 0 \\
                                   \end{array}
                                 \right) \ ,  \ \ \ L_L = \left(
                                                            \begin{array}{c}
                                                              \nu_L \\
                                                              e_L \\
                                                            \end{array}
                                                          \right)\ .
\eea
After expanding the ${SU(2)}$ indices it takes the form,
\bea
\mathcal{L} &=& -\lambda_u^{ij} \bar{u}_{\alpha R}^i(V_\alpha e_L^j - Y_\alpha \nu_L^j) \nonumber\\
&&-\lambda_e^{ij} \bar{e}_{R}^i(V_\alpha^\dagger u_{\alpha L}^j + Y_\alpha^\dagger d_{\alpha L}^j) + {\rm h.c.}\ .
\eea
Note that in this model the left-handed charged lepton fields couple to right-handed top quarks, and the right-handed charged lepton fields couple to left-handed top quarks. So a charged lepton chirality flip can be caused by the top mass at one loop.\\

\noindent $\textbf{Model II:\ \ \ }X = \left(3,2, 1/6\right)$. \\

The corresponding Lagrangian is,
\bea
\mathcal{L} &=& -\lambda_d^{ij} \bar{d}_{R}^i X^T\epsilon L_L^j + {\rm h.c.}\ ,
\eea
where we have used the same notation as in the previous case.
Expanding the ${SU(2)}$ indices yields,
\bea
\mathcal{L} &=& -\lambda_d^{ij} \bar{d}_{\alpha R}^i(V_\alpha e_L^j - Y_\alpha \nu_L^j)  + {\rm h.c.}\ .
\eea
In model II the leptoquark cannot couple to the top quark, so there is no $m_t$  enhancement in the $\mu \rightarrow e \gamma$ decay rate. There is also no $m_b$ enhancement, and the one-loop effective Hamiltonian for  $\mu \rightarrow e \gamma$ (after integrating out the massive scalars and the heavy quarks) is proportional to the muon mass.
For this reason, in the remainder of the paper we will focus entirely on model I.

\section{Phenomenology}
In this section we analyze some of the phenomenology  of model I, i.e., $X = \left(3,2, 7/6\right)$.
We concentrate only on those constraints which are most restrictive for the model and potentially most sensitive to the unusual top mass enhancement of the charged lepton chirality change, i.e., the ones coming from the following processes -- muon decay to an electron and a photon, muon to electron conversion, and electric dipole moment of the electron.

\subsection{Naturalness}
There is a logarithmically divergent contribution to the charged lepton mass matrix that  is enhanced by $m_t$. This contribution to the mass matrix, coming from momenta between $\Lambda $ (the cutoff) and $m_V$ is,
\bea\label{mij}
\Delta m_{i j} \simeq \tilde{\lambda}_u^{3i} \tilde{\lambda}_e^{j 3} \, \frac{3\,m_t}{16 \pi^2}  \log\!\left(\frac{\Lambda^2}{m_V^2}\right)\ .
\eea
To avoid unnatural cancellations between this loop contribution to the lepton mass matrix and the tree-level lepton mass matrix we require,
\bea\label{Delta_mij}
|\Delta m_{i j}| \lesssim \sqrt{m_i m_j}\ .
\eea
For example, for a scalar of mass $m_V = 50 \ \rm TeV$ and a cutoff set at the GUT scale Eq.~(\ref{mij}) gives,
\bea
\Delta m_{i j} \simeq \tilde{\lambda}_u^{3i} \tilde{\lambda}_e^{j 3}  \times 170 \ \rm GeV\ ,
\eea
which, combined with Eq.~(\ref{Delta_mij}), yields the following constraint on the couplings,
\bea
|\tilde{\lambda}_e^{13} \tilde{\lambda}_u^{32}| , |\tilde{\lambda}_e^{23} \tilde{\lambda}_u^{31}| \lesssim 4.3 \times 10^{-5}\ .
\eea
In the subsequent analysis we will include the constraint imposed by Eq.~(\ref{Delta_mij}) by indicating which regions of the plots are not favored by the naturalness considerations.

\subsection{$\mu \rightarrow e \gamma$ decay}
The relevant Feynman diagrams for this process are presented in Fig.~1.
\begin{figure}[t!]
\centering 
\includegraphics[trim=5cm 12.8cm 2.5cm 5.5cm, clip=true, totalheight=0.21\textheight]{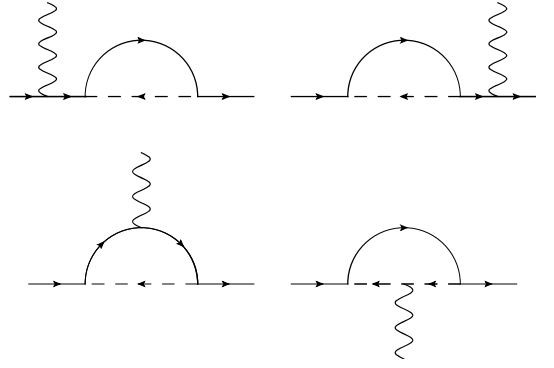}
\caption{Feynman diagrams contributing to the process $\mu \rightarrow e \gamma$.}
\label{fig1}
\end{figure}
The uniqueness of model I is that, apart from the fact there is no tree-level proton decay, the $\mu \rightarrow e \gamma$ rate is enhanced by the top quark mass.
To our knowledge, such an enhancement of $\mu \rightarrow e \gamma$ was observed previously only in \cite{Benbrik:2008si} in the context of an ${SU(2)}$ singlet scalar leptoquark model. However, that model suffers from perturbative proton decay and the impact of the $m_t$ enhancement was not focussed on.

Keeping only the piece enhanced by $m_t$, the sum of amplitudes corresponding to the diagrams in Fig.~1 (neglecting the terms proportional to $m_e$) is given by,
\begin{eqnarray}
i \mathcal{M} &=&  -\frac{3\,e \, m_t}{16\,\pi^2 m_V^2} f(m_t^2/m_V^2) \, k_{\nu} \, \epsilon_{\mu}(k) \nonumber\\
& & \times \Big[ \tilde{\lambda}_e^{13} \tilde{\lambda}_u^{32} \ \bar{e}_R(p-k)\,\sigma^{\mu\nu}\mu_L(p) \nonumber\\
& & +\,(\tilde{\lambda}_u^{31})^* (\tilde{\lambda}_e^{23})^* \ \bar{e}_L(p-k)\,\sigma^{\mu\nu}\mu_R(p)  \Big] \ ,
\end{eqnarray}
where $k$ is the photon four-momentum and $\epsilon$ is the photon polarization. The function $f(m_t^2/m_V^2)$ is given by,
\bea
f(x) = \frac{1-x^2+2 x \log x}{2(1-x)^3} + \frac{2}{3}\left(\frac{1-x+\log x}{(1-x)^2}\right) ,
\eea
and the tilde over the couplings denotes that they are related by transformations that take the quarks and leptons to their mass eigenstate basis through the following $3\times3$ matrix transformations,
\begin{equation}
{\tilde \lambda}_u= U(u,R)^{\dagger}\lambda_u U(e,L)\ , \ ~{\tilde \lambda}_e=U(e,R)^{\dagger} \lambda_e U(u,L)\ ,
\end{equation}
where the right-handed up quarks in the Lagrangian are related to the right-handed mass eigenstate up-type quarks by the matrix $U(u,R)$, the left-handed up quarks in the Lagrangian are related to the left-handed mass eigenstate up-type quarks by the matrix $U(u,L)$, {\it etc.}.

\begin{figure}[t!]
\centering 
\includegraphics[scale=1.255]{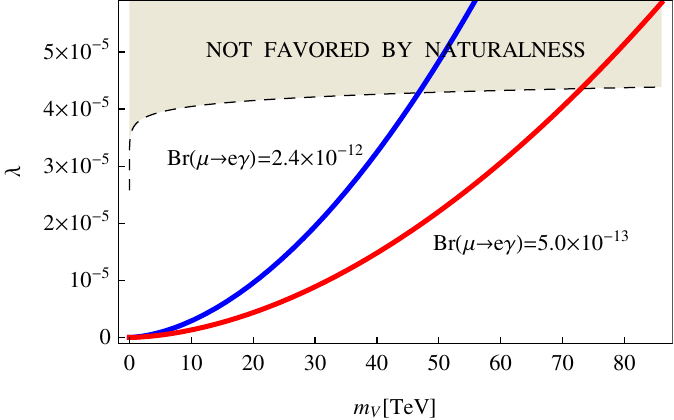}
\caption{The combination of couplings $\lambda$ from Eq.~(\ref{lambda}) as a function of the scalar leptoquark mass for two values of the $\mu\rightarrow e \gamma$ branching ratio relevant for the MEG experiment. The shaded region consists of points which do not satisfy Eq.~(\ref{Delta_mij}).}
\label{fig1}
\end{figure}

The $\mu \rightarrow e \gamma$ decay rate is,
\bea
{\Gamma}(\mu \rightarrow e \gamma) =\frac{9\,e^2 \lambda^2 m_t^2 m_\mu^3 }{2048\,\pi^5 m_V^4} f(m_t^2/m_V^2)^2\ ,
\eea
where,
\vspace{-4mm}
\bea\label{lambda}
\lambda \equiv \sqrt{\frac{1}{2}\big|\tilde{\lambda}_e^{13} \tilde{\lambda}_u^{32}\big|^2+ \frac{1}{2}\big|\tilde{\lambda}_u^{31} \tilde{\lambda}_e^{23}\big|^2}\ .
\eea
Fig.~2
shows the relation between $\lambda$ and the scalar leptoquark mass.
This dependence was plotted for the $\mu \rightarrow e \gamma$ branching ratio equal to the current upper limit of ${\rm Br}(\mu\rightarrow e \gamma) \simeq 2.4 \times 10^{-12}$ reported by the MEG experiment, and the prospective MEG sensitivity of ${\rm Br}(\mu\rightarrow e \gamma) \simeq 5.0 \times 10^{-13}$. It shows that the experiment will be sensitive to  scalar leptoquark masses at the hundred TeV scale for small values of the couplings.

For very small $x$, $f(x) \rightarrow \tilde{f}(x) = \frac{2}{3} \log{x}$.  This is a reasonable approximation in the range of $x$ we are interested in.  For example, $\tilde{f}(10^{-8})/f(10^{-8}) \simeq 1.1$.
\\
\vspace{2mm}

\subsection{$\mu \rightarrow e$ conversion}
\vspace{-2mm}
The effective Hamiltonian for the $\mu \rightarrow e$ conversion arises from two sources,
\begin{equation}
\mathcal{H}_{\rm eff}=\mathcal{H}_{\rm eff}^{\rm (a)}+\mathcal{H}_{\rm eff}^{\rm (b)}\ .
\end{equation}
The first is the dipole transition operator that comes from the loop diagrams which are responsible for the $\mu \rightarrow e \gamma$ decay, given by,
\begin{eqnarray}
\mathcal{H}_{\rm eff}^{\rm (a)}=\!\!\!\!&& {3\,e\, m_t \over 32 \pi^2 m_V^2}f(m_t^2/m_V^2)\left[{\tilde \lambda}_e^{13}{\tilde  \lambda}_u^{32}\,{\bar e}_R \sigma_{\mu \nu}\mu_L F^{\mu \nu} \right. \nonumber \\
&& + \left. ({\tilde \lambda}_u^{31})^*({ \tilde \lambda}_e^{23 })^* \,{\bar e}_L \sigma_{\mu \nu} \mu_R  F^{\mu \nu} \right]\ .
\end{eqnarray}

Using the following Fierz identities (for spinors),
\bea
(\bar{u}_{1L} u_{2R})(\bar{u}_{3R} u_{4L}) &=& \frac{1}{2}(\bar{u}_{1L} \gamma^\mu u_{4L})(\bar{u}_{3R} \gamma_\mu u_{2R})\ ,\nonumber\\
(\bar{u}_{1L} u_{2R})(\bar{u}_{3L} u_{4R}) &=& \frac{1}{2}(\bar{u}_{1L} u_{4R})(\bar{u}_{3L} u_{2R})\\
&+& \frac{1}{8}(\bar{u}_{1L} \sigma^{\mu\nu} u_{4R})(\bar{u}_{3R} \sigma_{\mu\nu} u_{2L})\,,\nonumber
\eea
we arrive, after integrating out the heavy scalar leptoquarks (at tree level), at the second  part of the effective Hamiltonian,
\begin{widetext}
\bea\label{Hamiltonian}
\mathcal{H}_{\rm eff}^{\rm (b)} &=& \frac{1}{2 m_V^2} \bigg\{\tilde{\lambda}_u^{12}(\tilde{\lambda}_u^{11})^*(\bar{e}_{L} \gamma^\mu \mu_{L})(\bar{u}_{\alpha R} \gamma_\mu u_{\alpha R})+\tilde{\lambda}_e^{11}\tilde{\lambda}_u^{12}\Big[C_S(\mu)(\bar{e}_{R}  \mu_{L})(\bar{u}_{\alpha R}  u_{\alpha L})+\frac{1}{4}C_T(\mu)(\bar{e}_{R} \sigma^{\mu\nu} \mu_{L})(\bar{u}_{\alpha R} \sigma_{\mu \nu} u_{\alpha L})\Big]\nonumber\\
& &+ \ \tilde{\lambda}_e^{11}(\tilde{\lambda}_e^{21})^*(\bar{e}_{R} \gamma^\mu \mu_{R})(\bar{u}_{\alpha L} \gamma_\mu u_{\alpha L})\!+\!(\tilde{\lambda}_e^{21})^*(\tilde{\lambda}_u^{11})^*\!\Big[C_S(\mu)(\bar{e}_{L}  \mu_{R})(\bar{u}_{\alpha L}  u_{\alpha R})\!+\!\frac{1}{4}C_T(\mu)(\bar{e}_{L} \sigma^{\mu\nu} \mu_{R})(\bar{u}_{\alpha L} \sigma_{\mu \nu} u_{\alpha R})\Big]\!\bigg\}\nonumber\\
& & + \ \frac{1}{2 m_Y^2}(\tilde{\lambda}_e V_{CKM})^{11}\left((\tilde{\lambda}_e V_{CKM})^{21}\right)^*(\bar{e}_{R} \gamma^\mu \mu_{R})(\bar{d}_{\alpha L} \gamma_\mu d_{\alpha L})\ +\ldots \ .
\eea
\end{widetext}
The CKM matrix arises whenever a coupling to the left-handed down-type quark appears. In Eq.~(\ref{Hamiltonian}) the contribution of the heavy quarks, as well as the contribution of the strange quark, are in the ellipses.  Since the operators $\bar q q$ and $\bar q \sigma^{\mu \nu}q$ do require renormalization, their matrix elements develop subtraction point dependence that is cancelled in the leading logarithmic approximation by that of the coefficients $C_{S,T}$. Including strong interaction leading logarithms we get,
\begin{equation}
C_S(\mu)=\left[\alpha_s(m_V) \over \alpha_s(\mu) \right]^{-12/( 33-2N_q)}
\end{equation}
and
\begin{equation}
C_T(\mu)=\left[\alpha_s(m_V) \over \alpha_s(\mu) \right]^{4/( 33-2N_q)}\ ,
\end{equation}
where $N_q=6$ is the number of quarks with mass below $m_V$.
In order to match the effective Hamiltonian (\ref{Hamiltonian}) to the Hamiltonian at the nucleon level and use this to compute the conversion rate, we follow the steps outlined in \cite{Kitano:2002mt,Cirigliano:2009bz}.

\begin{figure}[t!]
\centering 
\includegraphics[scale=1.24]{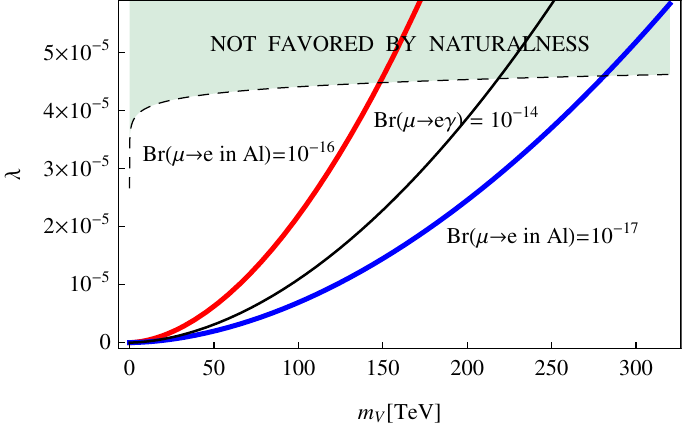}
\caption{The combination of couplings $\lambda$ from Eq.~(\ref{lambda}) as a function of the scalar leptoquark mass for two values of the ${\rm Br}(\mu\rightarrow e \ {\rm conversion\ in\ Al})$ relevant for the Mu2e experiment.  The thin solid line, corresponding to ${\rm Br}(\mu\rightarrow e \gamma) = 10^{-14}$, is included for reference. The shaded region consists of points which do not satisfy Eq.~(\ref{Delta_mij}).}
\label{fig1}
\end{figure}

Our results, taking into account only the contribution from  $\mathcal{H}_{\rm eff}^{\rm (a)} $, are shown in Fig.~3.
The current experimental limit is ${\rm Br}(\mu\rightarrow e \ {\rm conversion\ in\ Au}) < 7.0 \times 10^{-13}$ \cite{Wintz:1998rp}.
However, here we focus on the prospective Mu2e experiment \cite{Abrams:2012er}, which has a sensitivity goal of $5 \times 10^{-17}$.  The COMET experiment \cite{Kuno} aims for comparable sensitivity in later stages.
We use the total capture rate for $_{13}^{27}\rm Al$ of $\omega_{\rm capture} = 0.7054 \times 10^6 \ {\rm {s^{-1}}}$ \cite{Suzuki:1987jf} to switch from the $\mu \rightarrow e$ conversion rate to a branching ratio.

Apart from coupling constant factors, the contribution to the $\mu \rightarrow e$ conversion amplitude from  $\mathcal{H}_{\rm eff}^{\rm (a)} $ is enhanced over the contribution to the amplitude from  $\mathcal{H}_{\rm eff}^{\rm (b)} $ roughly by $(m_t/m_{\mu})(3 e^2/32 \pi^2)\,{\rm log}(m_V^2/m_t^2) \sim  10$, for $m_V$ in the hundred TeV range.

Our results show that in some regions of parameter space the Mu2e experiment will be able to constrain leptoquark couplings with similar precision to what can be done with an experiment which is sensitive to a branching ratio for $\mu \rightarrow e \gamma$ of around $10^{-14}$. In other regions the Mu2e experiment is likely to give a  more powerful constraint for such a $\mu \rightarrow e \gamma$ branching ratio, for example, when the Yukawa couplings are strongly hierarchical and the top quark loop is very suppressed.

To show graphically the contributions to the branching ratio originating from terms in the effective Hamiltonian with different structures, we set all the couplings to zero apart from $\tilde{\lambda}_e^{13}, \tilde{\lambda}_e^{23}, \tilde{\lambda}_u^{31}, \tilde{\lambda}_u^{32}, \tilde{\lambda}_u^{11}, \tilde{\lambda}_u^{12}$ for simplicity, i.e., we leave only the couplings relevant for the $\mu \rightarrow e \gamma$ decay and one of the vector contributions to $\mathcal{H}_{\rm eff}^{\rm (b)}$.

\begin{figure}[t!]
\centering 
\includegraphics[scale=1.25]{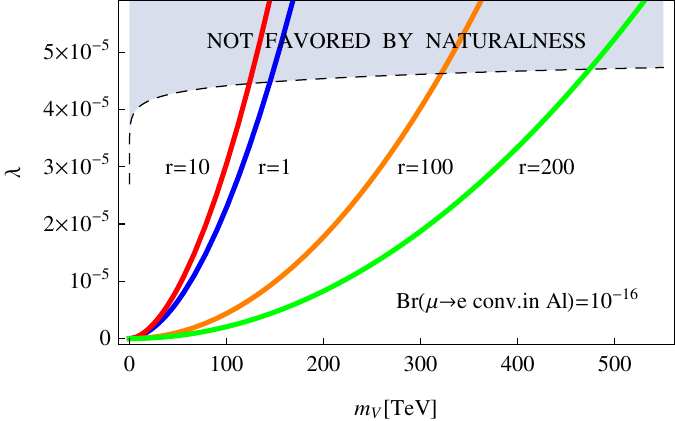}
\caption{The combination of couplings $\lambda$ from Eq.~(\ref{lambda}) as a function of the scalar leptoquark mass for a branching ratio ${\rm Br}(\mu\rightarrow e \ {\rm conversion\ in\ Al}) = 10^{-16}$ and four different positive values of the ratio of the couplings $r$ from Eq.~(\ref{ratio}). The shaded region consists of points which do not satisfy Eq.~(\ref{Delta_mij}).}
\label{fig1}
\end{figure}
\begin{figure}[t!]
\centering 
\includegraphics[scale=1.25]{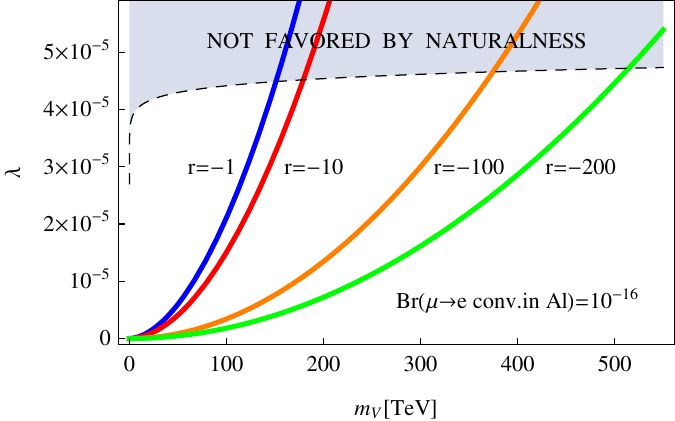}
\caption{Same as Fig.~4, but for negative values of $r$.}
\label{fig1}
\end{figure}

Note that the heavy quark contributions are suppressed by $\Lambda_{\rm QCD}/m_Q$, low energy phenomenology suggests that the strange quark contribution is small, and furthermore the tensor contributions are not enhanced by the atomic number of the target.

In addition, we consider only real couplings and define  $\kappa \equiv \tilde{\lambda}_u^{11} \tilde{\lambda}_u^{12}$. We also assume $\tilde{\lambda}_e^{13} \tilde{\lambda}_u^{32} = \tilde{\lambda}_u^{31} \tilde{\lambda}_e^{23} = \lambda$, so that we can plot $\lambda$  as a function of the scalar leptoquark mass $m_V$ for a given value of the ratio,
\bea\label{ratio}
r \equiv \frac{\kappa}{\lambda} = \frac{ \tilde{\lambda}_u^{11} \tilde{\lambda}_u^{12}}{\sqrt{\frac{1}{2}(\tilde{\lambda}_e^{13} \tilde{\lambda}_u^{32})^2+ \frac{1}{2}(\tilde{\lambda}_u^{31} \tilde{\lambda}_e^{23})^2}}\ .
\eea
Figs.~4 -- 7 show our results for a few values of $r = $ $\pm 1$, $\pm 10$, $\pm 100$, $\pm 200$ and two values of the branching ratio ${\rm Br}(\mu\rightarrow e \ {\rm conversion\ in\ Al}) = 10^{-16}, 10^{-17}$.

For $r \lesssim 1$ the branching ratio is dominated by the $\mathcal{H}_{\rm eff}^{\rm (a)} $ contribution and in this parameter region all curves look like the ones in Fig.~3. For larger values of $r$, depending on the relative sign between the contributions from $\mathcal{H}_{\rm eff}^{\rm (a)} $ and $\mathcal{H}_{\rm eff}^{\rm (b)} $, there are two possibilities. If the interference is constructive, the curve moves down with increasing $r$ since a smaller value of the coupling $\lambda$ is required to achieve a given branching ratio (Figs.~5, 7). In the case of a destructive interference, the curves move up until a value of $r$ is reached for which the two contributions are the same (Figs.~4, 6). As estimated before, this occurs for $r \approx 10$. Increasing $r$ further brings the curves back down, since the  $\mathcal{H}_{\rm eff}^{\rm (b)} $ contribution becomes dominant.

Large values of $r$ are expected if the Yukawa couplings of $X$ exhibit a hierarchical pattern like what is observed in the quark sector; $\kappa$ changes generations by one unit while the product of couplings in $\lambda$ involves changing generations by three units.
Finally, we note that for all the curves in the plots above the Yukawa couplings are well within the perturbative regime.

\begin{figure}[t!]
\centering 
\includegraphics[scale=1.25]{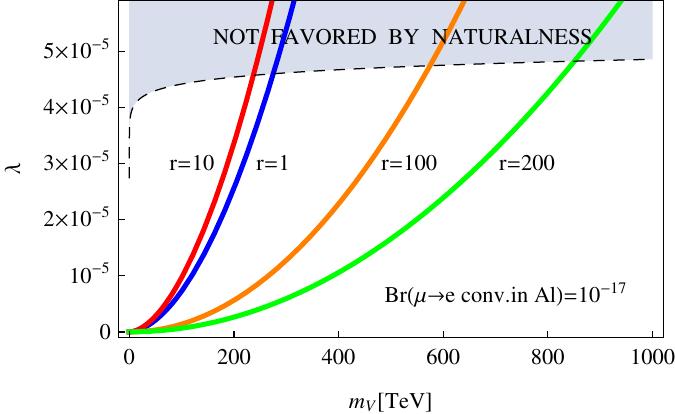}
\caption{Same as Fig.~4, but for a branching ratio ${\rm Br}(\mu\rightarrow e \ {\rm conversion\ in\ Al}) = 10^{-17}$.}
\label{fig1}
\end{figure}
\begin{figure}[t!]
\centering 
\includegraphics[scale=1.25]{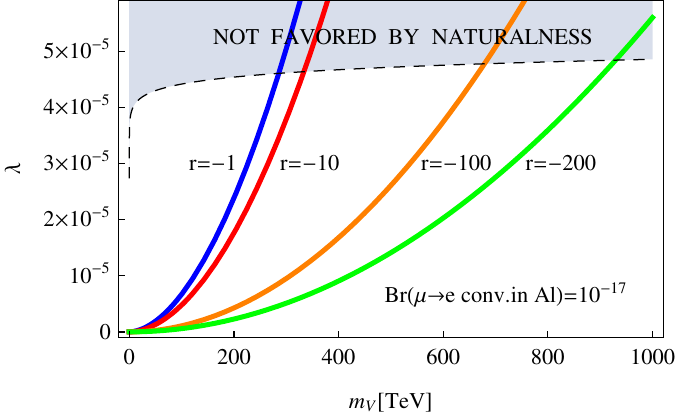}
\caption{Same as Fig.~5, but for a branching ratio ${\rm Br}(\mu\rightarrow e \ {\rm conversion\ in\ Al}) = 10^{-17}$.}
\label{fig1}
\end{figure}

\subsection{Electron EDM}
Another  flavor constraint on the couplings of model I comes from the electric dipole moment (EDM) of the electron.
As mentioned earlier, the fact that $X$ couples directly to both left- and right-handed quarks means that at one loop the top quark mass can induce the chirality flip necessary to give an electron EDM. We find that,
\bea\label{edm}
\left|d_e\right| \simeq \frac{3\,e\, m_t}{16\,\pi^2 m_V^2} f(m_t^2/m_V^2) \big|{\rm Im} [\tilde{\lambda}_e^{13} \tilde{\lambda}_u^{31}]\big|\ .
\eea
The present electron EDM experimental limit \cite{Hudson:2011zz} is,
\bea
|d_e| < 10.5 \times 10^{-28} \ e \ \rm cm\ .
\eea
We can write the dipole moment in terms of the branching ratio, ${\rm Br}(\mu \rightarrow e \gamma)$, giving the constraint,
\begin{align}
\frac{\big| {\rm Im} [\tilde{\lambda}_e^{13} \tilde{\lambda}_u^{31}]\big|}{\lambda} \sqrt{{\rm Br}(\mu \rightarrow e \gamma)} < 2.0 \times 10^{-7} \ .
\end{align}
For example, if model I gave a branching ratio equal to the current experimental bound of ${\rm Br}(\mu \rightarrow e \gamma) < 2.4 \times 10^{-12}$, this would correspond to the constraint on the couplings of $\big| {\rm Im} [\tilde{\lambda}_e^{13} \tilde{\lambda}_u^{31}]\big|/ \lambda < 0.13$.
Figure 8 shows the relation between the parameters $\big| {\rm Im} [\tilde{\lambda}_e^{13} \tilde{\lambda}_u^{31}]\big|$ and $m_V$ for the electron EDM equal to $|d_e| = 10^{-27}, \ 10^{-28}, \ {\rm and \ } 10^{-29} \ e \ \rm cm.$

\begin{figure}[t!]
\centering
\includegraphics[scale=1.24]{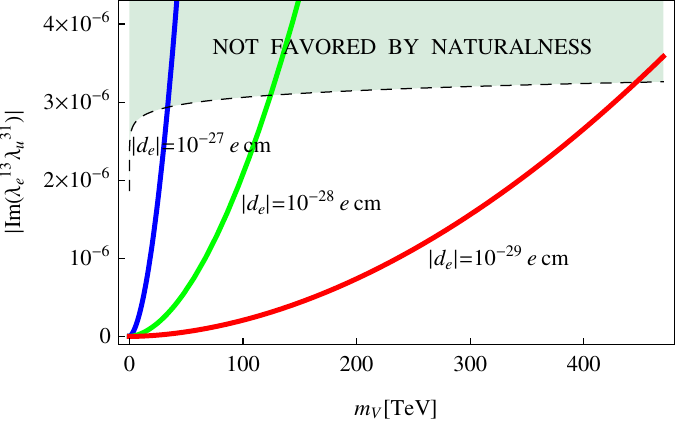}
\caption{The combination of couplings $\big| {\rm Im} [\tilde{\lambda}_e^{13} \tilde{\lambda}_u^{31}]\big|$  as a function of the scalar leptoquark mass for three different values of the electron EDM. The shaded region consists of points which do not satisfy Eq.~(\ref{Delta_mij}).}
\label{fig1}
\end{figure}

\section{Baryon number violation and dimension five operators}
Tree-level renormalizable interactions are not the only possible source of baryon number violation. It might also occur through higher-dimensional nonrenormalizable operators. In the standard model proton decay is restricted to operators of mass dimension six or higher. However, the scalar leptoquark models we consider exhibit proton decay through dimension five operators.

Let's first consider model I, in which $X = \left(3,2, 7/6\right)$. Although it doesn't give proton decay at tree level, one can construct the following dimension five operator,
\bea\label{O1}
\mathcal{O}_{I} = \frac{1}{\Lambda} \, g^{a b} d_{R \alpha}^a d_{R \beta}^b (H^\dagger X_{\gamma}) \epsilon^{\alpha \beta \gamma}\ .
\eea
The coupling constant matrix $g$ is antisymmetric in flavor space. Because of the tree-level leptoquark couplings (see, Table I), baryon number violating decay occurs here through the process shown in Fig.~9, resulting in $n \rightarrow e^- K^+$ and $p \rightarrow  K^+ \nu$.
\begin{figure}[t!]
\centering 
\includegraphics[scale=0.6]{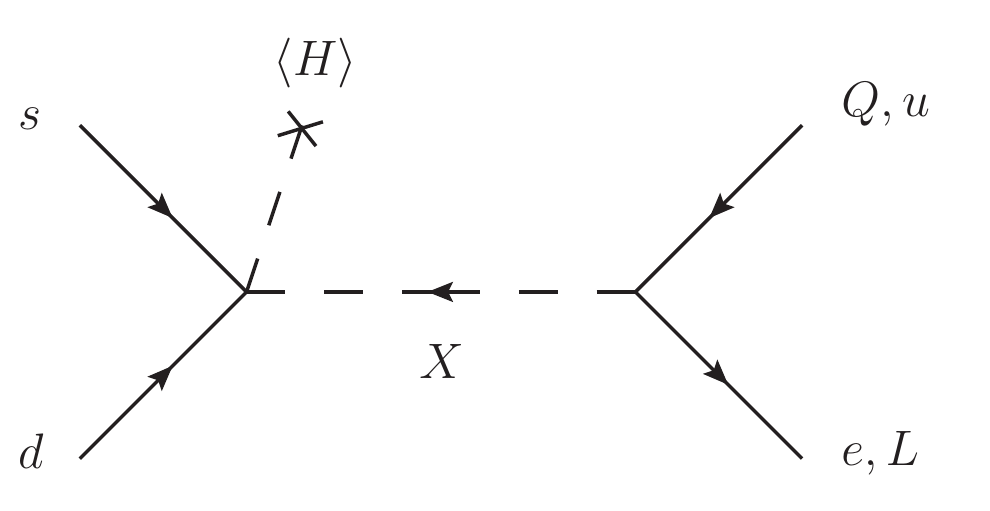}
\caption{Feynman diagram representing proton decay in model I.}
\label{fig1}
\end{figure}
Setting the coupling constants to unity, we estimate the baryon number violating nucleon decay rate caused by this operator to be,
\bea\label{rate}
\Gamma_p \approx 2\times 10^{-57}\left(\frac{50 \ \rm TeV}{m_V}\right)^4\left(\frac{M_{\rm PL}}{\Lambda}\right)^2 {\rm GeV}\ .
\eea
Since the current experimental limit is $\Gamma_p^{\rm exp} < 2.7 \times 10^{-66} \ {\rm GeV}$ \cite{Nishino:2009aa}, even if the scale of new physics $\Lambda$ is equal to the Planck mass $M_{\rm PL}$ when the coupling constants are unity, this operator causes too large a proton decay rate for $m_V\lesssim10\,000~{\rm TeV}$.

In the case of model II, where $X = \left(3,2, 1/6\right)$, there are two dimension five baryon number violating operators,
\bea\label{O2}
\mathcal{O}_{II}^{(1)} &=& \frac{1}{\Lambda} \, g^{a b} u_{R \alpha}^a d_{R \beta}^b (H^\dagger X_{\gamma}) \epsilon^{\alpha \beta \gamma}\ ,\nonumber\\
\mathcal{O}_{II}^{(2)} &=&  \frac{1}{\Lambda} \, g^{a b} u_{R \alpha}^a e_{R}^b (X_{\beta}\epsilon X_\gamma) \epsilon^{\alpha \beta \gamma}\ .
\eea
The operator $\mathcal{O}_{II}^{(1)}$ permits a nucleon decay pattern similar to the previous case, e.g., $n \rightarrow e^- \pi^+$ and $p \rightarrow  \pi^+ \nu$.
Proton decay through the operator $\mathcal{O}_{II}^{(2)}$ is much more suppressed.

In order to prevent proton decay through dimension five operators, one could introduce a discrete gauge symmetry that  forbids the baryon number violating nonrenormalizable couplings. Since $B-L$ is the only anomaly free global symmetry in the standard model, we chose to impose a discrete subgroup of $B-L$.  In models I and II the leptoquark has $B-L=4/3$. The usual $Z_2$, where the nontrivial transformation is $(-1)^{B-L}$, doesn't work, as the operators $\mathcal{O}_I$, $\mathcal{O}_{II}^{(1)}$, and $\mathcal{O}_{II}^{(2)}$ are invariant under this transformation. However, we find that imposing a $Z_3$ discrete symmetry, with elements that are powers of  ${\rm exp}[2\pi i(B-L)/3]$, forbids these dimension five operators and, thus, prevents the proton from decaying in this class of models. Note that gauging $B-L$ and spontaneously breaking the symmetry with a charge three scalar (at some high scale) leaves this unbroken discrete $Z_3$ gauge symmetry. It is not possible to use any discrete subgroup of $B-L$ to forbid proton decays in the models from Table I which exhibit proton decay at tree level since all the interactions conserve $B-L$.

Finally, we would like to comment on the relation between this work and that of  \cite{Arnold:2012sd}, where renormalizable models that have  additional scalars  and have baryon number violation at tree level but not proton decay were enumerated and discussed. In these models none of the scalars were leptoquarks (they could rather be called diquarks or dileptons). However, if we permit higher dimension operators, then models 4 and 9 containing the scalar $X = \left(3, 1, 2/3\right)$ (which has renormalizable diquark couplings), have dimension five leptoquark-type couplings,
\bea
\mathcal{O}_{III} &=& \frac{1}{\Lambda} \, g^{a b} (\bar{Q}_{L}^{\alpha a} H) e_{R}^b X_{\alpha} \ .
\eea
This operator, combined with the renormalizable couplings of $X$ to two quarks, gives proton decay with the rate estimated in Eq. (\ref{rate}). This observation restricts the parameter space of models 4 and 9 presented in \cite{Arnold:2012sd} to the one in which either the color triplet scalar $X$ is very heavy or its Yukawa couplings are small.

\section{Conclusions}
We have investigated the minimal set of renormalizable models in which a single scalar leptoquark is added to the standard model with the requirement that proton decay not be induced in perturbation theory. We have looked in detail at one particular model which gives an unusual top quark mass enhancement of the branching ratio of $\mu \rightarrow e \gamma$.

For this model, we have considered the $\mu \rightarrow e \gamma$ branching ratio, the $\mu \rightarrow e$ conversion rate, and the electric dipole moment of the electron, in light of current constraints and future experiments.  We have shown the potential limits the MEG, Mu2e, and the electron EDM experiments could place on some of the couplings of the scalar leptoquark to the $\bar{Q} e$ and $L \bar{u}$ bilinears. We have explored the region of parameter space for which the loop contribution to the charged lepton mass matrix does not overwhelm the tree-level part. Given this naturalness constraint, we have found that current experiments are sensitive to leptoquark masses on the order of a hundred TeV, whereas future experiments may push the sensitivity into the several hundred TeV mass region.

We have commented on the existence of nonrenormalizable operators in these minimal models which can give an unacceptably large proton decay rate for $m_V \lesssim 10\,000 {\rm ~TeV}$, as well as provided a simple mechanism for avoiding them.

Since there are only two scalar leptoquark models where at the renormalizable level baryon number is automatically conserved, it would be interesting to examine a more extensive range of phenomena and address,  over a wide range of parameter space, how to distinguish experimentally between these two models.

\subsection*{Acknowledgment}
The authors are grateful to Vincenzo Cirigliano and Robert Tschirhart for helpful comments.
This paper is funded by the Gordon and Betty Moore Foundation through Grant $\# 776$ to the Caltech Moore Center for Theoretical Cosmology and Physics. The work of the authors was supported also in part by the U.S. Department of Energy under contract No. DE-FG02-92ER40701.



\begin{thebibliography}{99}




\bibitem{Adam:2011ch}
  J.~Adam {\it et al.}  [MEG Collaboration],
  \textit{New limit on the lepton-flavour violating decay $\mu^{+} \to e^{+} \gamma$},
  Phys.\ Rev.\ Lett.\  {\bf 107}, 171801 (2011)
  [arXiv:1107.5547 [hep-ex]].

\bibitem{Abrams:2012er}
  R.~J.~Abrams {\it et al.}  [Mu2e Collaboration],
  \textit{Mu2e conceptual design report},
  arXiv:1211.7019 [physics.ins-det].


\bibitem{Kuno}
  Y.~Kuno {\it et al.}  [COMET Collaboration],
  \textit{A search for muon-to-electron conversion at J-PARC: the COMET experiment},
  Prog.\ Theor.\ Exp.\ Phys.\ {\bf 2013}, 022C01.


\bibitem{Buchmuller:1986zs}
  W.~Buchmuller, R.~Ruckl and D.~Wyler,
  \textit{Leptoquarks in lepton - quark collisions},
  Phys.\ Lett.\ B {\bf 191}, 442 (1987)
  [Erratum-ibid.\ B {\bf 448}, 320 (1999)].


\bibitem{Davies:1990sc}
  A.~J.~Davies and X.~-G.~He,
  \textit{Tree level scalar fermion interactions consistent with the symmetries of the standard model},
  Phys.\ Rev.\ D {\bf 43}, 225 (1991).


\bibitem{Davidson:1993qk}
  S.~Davidson, D.~C.~Bailey and B.~A.~Campbell,
  \textit{Model independent constraints on leptoquarks from rare processes},
  Z.\ Phys.\ C {\bf 61}, 613 (1994)
  [hep-ph/9309310].



\bibitem{Gabrielli:2000te}
  E.~Gabrielli,
  \textit{Model independent constraints on leptoquarks from rare $\mu$ and $\tau$ lepton processes},''
  Phys.\ Rev.\ D {\bf 62}, 055009 (2000)
  [hep-ph/9911539].


\bibitem{Benbrik:2008si}
  R.~Benbrik and C.~-K.~Chua,
  \textit{Lepton flavor violating $l \rightarrow l' \gamma$ and $Z \rightarrow l \bar{l}'$ decays induced by scalar leptoquarks},
  Phys.\ Rev.\ D {\bf 78}, 075025 (2008)
  [arXiv:0807.4240 [hep-ph]].


\bibitem{Dorsner:2009cu}
  I.~Dorsner, S.~Fajfer, J.~F.~Kamenik and N.~Kosnik,
  \textit{Can scalar leptoquarks explain the $f_{D_s}$ puzzle?},
  Phys.\ Lett.\ B {\bf 682}, 67 (2009)
  [arXiv:0906.5585 [hep-ph]].

\vspace{0.7mm}


\bibitem{Benbrik:2010cf}
  R.~Benbrik, M.~Chabab and G.~Faisel,
  \textit{Lepton flavour violating $\tau$ and $\mu$ decays induced by scalar leptoquark},
  arXiv:1009.3886 [hep-ph].

  \bibitem{Gonderinger:2010yn}
  M.~Gonderinger and M.~J.~Ramsey-Musolf,
  \textit{Electron-to-tau lepton flavor violation at the electron-ion collider},
  JHEP {\bf 1011}, 045 (2010)
  [Erratum-ibid.\  {\bf 1205}, 047 (2012)]
  [arXiv:1006.5063 [hep-ph]].


\bibitem{Arnold:2012sd}
  J.~M.~Arnold, B.~Fornal and M.~B.~Wise,
  \textit{Simplified models with baryon number violation but no proton decay},
  Phys.\  Rev.\ D {\bf 87}, 075004 (2013)
  [arXiv:1212.4556 [hep-ph]].



\bibitem{Kitano:2002mt}
  R.~Kitano, M.~Koike and Y.~Okada,
  \textit{Detailed calculation of lepton flavor violating muon electron conversion rate for various nuclei},
  Phys.\ Rev.\ D {\bf 66}, 096002 (2002)
  [Erratum-ibid.\ D {\bf 76}, 059902 (2007)]
  [hep-ph/0203110].


\bibitem{Cirigliano:2009bz}
  V.~Cirigliano, R.~Kitano, Y.~Okada and P.~Tuzon,
  \textit{On the model discriminating power of $\mu \rightarrow e$ conversion in nuclei},
  Phys.\ Rev.\ D {\bf 80}, 013002 (2009)
  [arXiv:0904.0957 [hep-ph]].





\bibitem{Wintz:1998rp}
  P.~Wintz, [for the SINDRUM Collaboration]
  \textit{Results of the SINDRUM-II experiment},
  Conf.\ Proc.\ C {\bf 980420}, 534 (1998).



\bibitem{Suzuki:1987jf}
  T.~Suzuki, D.~F.~Measday and J.~P.~Roalsvig,
  \textit{Total nuclear capture rates for negative muons},
  Phys.\ Rev.\ C {\bf 35}, 2212 (1987).



\bibitem{Hudson:2011zz}
  J.~J.~Hudson, D.~M.~Kara, I.~J.~Smallman, B.~E.~Sauer, M.~R.~Tarbutt and E.~A.~Hinds,
  \textit{Improved measurement of the shape of the electron},
  Nature {\bf 473}, 493 (2011).



\bibitem{Nishino:2009aa}
  H.~Nishino {\it et al.}  [Super-Kamiokande Collaboration],
  \textit{Search for proton decay via $p \rightarrow e^+ \pi^0$ and $p \rightarrow \mu^+ \pi^0$ in a Large Water Cherenkov Detector},
  Phys.\ Rev.\ Lett.\  {\bf 102}, 141801 (2009)
  [arXiv:0903.0676 [hep-ex]].

\end{thebibliography}
\end{document}